\documentclass[a4,11pt]{article}

\usepackage{graphicx}
\usepackage[a4paper, left=3cm, right=3cm, top=3cm, bottom=3cm]{geometry}
\usepackage{bm}
\usepackage{amsmath,amsfonts,amssymb}
\usepackage{lipsum}

\newcommand{\transpose}[1]{#1^\mathrm{T}}
\newcommand{\ith}[1]{#1^\text{th}}
\newcommand{\grpT}[1]{\bm{#1_\mathrm{T}}}
\newcommand{\grpA}[1]{\bm{#1_\mathrm{A}}}
\newcommand{\grpB}[1]{\bm{#1_\mathrm{B}}}
\newcommand{\XA}{\grpA{X}}
\newcommand{\XB}{\grpB{X}}
\newcommand{\XC}{\bm{X_\mathrm{C}}}
\newcommand{\XS}{\bm{X_\mathrm{S}}}
\newcommand{\XSK}{\bm{X_{\mathrm{S}_K}}}

\newcommand{\XCA}{\bm{X_\mathrm{CA}}}
\newcommand{\XCB}{\bm{X_\mathrm{CB}}}

\newcommand{\oneT}[1]{\bm{{#1}^\mathrm{T}_{(1)}}}
\newcommand{\twoT}[1]{\bm{{#1}^\mathrm{T}_{(2)}}}
\newcommand{\betaV}{\bm{\beta}}
\newcommand{\betaA}{\grpA{\beta}}
\newcommand{\betaAK}{\bm{\beta_{\mathrm{A}_K}}}
\newcommand{\betaB}{\grpB{\beta}}
\newcommand{\betaBK}{\bm{\beta_{\mathrm{B}_K}}}
\newcommand{\betaC}{\bm{\beta_\mathrm{C}}}

\newcommand{\betaVH}{\bm{\hat{\beta}}}
\newcommand{\betaAH}{\grpA{\hat{\beta}}}
\newcommand{\betaBH}{\grpB{\hat{\beta}}}
\newcommand{\X}{\bm{X}}
\newcommand{\ba}{\bm{\alpha}}
\newcommand{\one}[1]{\bm{{#1}_{(1)}}}

\newcommand{\two}[1]{\bm{{#1}_{(2)}}}

\newcommand{\XT}{\bm{X^\mathrm{T}}}
\newcommand{\y}{\bm{y}}
\newcommand{\yA}{\grpA{y}}
\newcommand{\yB}{\grpB{y}}
\newcommand{\rank}[1]{\bm{#1^\mathrm{(r)}}}

\newcommand{\Cov}[2]{\text{Cov}\left(#1\text{, }#2\right)}
\newcommand{\gammaV}{\bm{\gamma}}
\newcommand{\gammaVH}{\bm{\hat{\gamma}}}

\newcommand{\WH}{\bm{\widehat{W}}}
\newcommand{\Wt}{\bm{\widetilde{W}}}
\newcommand{\z}{\bm{z}}
\newcommand{\zt}{\bm{\widetilde{z}}}

\newcommand{\model}[1]{\mathcal{M}_\mathrm{#1}}
\newcommand{\E}[1]{\mathbb{E}\left[#1\right] }
\newcommand{\MA}{\model{A}}

\newcommand{\std}[1]{{#1}^{(\zeta)}}

\newcommand{\oxStats}{\text{Department of Statistics, University of Oxford, UK}}
\newcommand{\oxNDM}{\text{Nuffield Department of Medicine, University of Oxford, UK}}
\newcommand{\ATI}{\text{Alan Turing Institute, London, UK}}
\newtheorem{subgTh}{Theorem}

\title{Supervised variable selection in randomised controlled trials prior to exploration of treatment effect heterogeneity: an example from severe malaria}
\author{Chieh-Hsi Wu\thanks{\oxStats} and Chris C. Holmes\footnotemark[1] \thanks{\oxNDM}\thanks{\ATI}}

\begin{document}
\maketitle

\begin{abstract}
Exploration of treatment effect heterogeneity (TEH) is an increasingly important aspect of modern statistical analysis for stratified medicine in randomised controlled trials (RCTs) as we start to gather more information on trial participants and wish to maximise the opportunities for learning from data.
However, the analyst should refrain from including a large number of variables in a treatment interaction discovery stage.
Because doing so can significantly dilute the power to detect any true outcome-predictive interactions between treatments and covariates.
Current guidance is limited and mainly relies on the use of unsupervised learning methods, such as hierarchical clustering or principal components analysis, to reduce the dimension of the variable space prior to interaction tests.
In this article we show that outcome-driven dimension reduction, i.e. supervised variable selection, can maintain power without inflating the type-I error or false-positive rate.
We provide theoretical and applied results to support our approach. The applied results are obtained from illustrating our framework on the dataset from an RCT in severe malaria.
We also pay particular attention to the internal risk model approach for TEH discovery which we show is a particular case of our method and we point to improvements over current implementation.
\end{abstract}

\section{Introduction \label{section intro}}
Randomised control trials (RCTs) are considered to be the most reliable means of comparing the effects of multiple treatments or health interventions, sometimes including a control group (placebo or no treatment).
It is increasingly common to exploit the opportunities to conduct retrospective treatment interaction analysis (TIA)\footnote{Such analyses also go under the name of treatment effect heterogeneity (TEH) or treatment effect modifiers (TEM).} on predictor variables in RCT data \cite{altman:2015cl}.
In such cases, the expected effect of a treatment is dependent on the value of one or more predictor variables, for example if the effect of a treatment varies with the age of the subject.
Improvement in the volume of measurements obtainable from each patient (e.g. genomic profiling) leads to a myriad of candidate factors for interaction analysis.
Given the large costs associated with clinical trials, there is a natural desire to learn from the data as much as possible.
However, statistical power is not free and any treatment interaction analysis is certainly not immune from false positive rates.
These are the problems that underlay many challenges of performing a valid interaction analysis.  
Our motivating example arises from a severe malaria trial, SEAQUAMAT \cite{dondorp:2005ar}, where investigating the treatment effect heterogeneity can provide valuable insight on the mechanism of the disease.
Moreover, if the utility of a treatment composes of other factors (e.g. costs, side effects etc) additional to the treatment effect, then information on treatment effect heterogeneity can help us make better decisions on medical resource management.

Clearly the power to detect an effect modifier is less than that to detect an overall treatment effect for an equivalent effect size \cite{brookes:2001su,brookes:2004su,rothwell:2005su,burke:2015th}.
In addition, for most interaction analyses the original trial will have been designed to detect an overall main treatment effect without tests for interactions taken into consideration \cite{dondorp:2005so,foster:2015ef,seymour:2007di}. 
This can contribute to a lack of power for discovery unless the interaction is larger in magnitude than that assumed for the main effect in the study design \cite{altman:2015cl}, as for subgroups that benefit from a change in treatment the relative population size is smaller by definition.

If a large number of interaction tests are conducted, the study can also suffer from spurious false positives arising from multiple comparisons \cite{ioannidis:2005mi,kent:2010as,burke:2015th}.
On the other hand, false negatives arise due to inadequate power to withstand correction for multiple testing \cite{kent:2010as,burke:2015th}.
If the measurement covariates are dependent, further power can be lost when testing each variable interaction one at a time rather than in a joint analysis \cite{rothwell:2005su,kent:2010as}.

For the reasons above, several recommendations have been made on such analyses to increase the reliability of their findings and better investment of statistical power \cite{rothwell:2005su,kent:2010as,sun:2014us,burke:2015th}.
There is a strong emphasis on a pre-specified analysis plan, wherein the hypotheses are defined \textit{a priori}.
Of these hypotheses, the primary analyses must be distinguished from the secondary analyses and several authors suggest that the primary hypotheses be restricted to a small number (one to two analyses) supported by external theoretical and empirical evidence.

These recommendations send a common message that an interaction analysis plan should be formulated with care.
Careful expenditure of power also resonates with the rationale for using a single global test for interaction discovery, which provides a single p-value against the null hypothesis of ``no treatment interactions.'' 
It can be envisaged that such a test would require tools that can effectively and efficiently learn about the structure of and the patterns in the entire dataset, which is a challenging task when the dimension the variable space is high.
Machine learning (ML) and AI methods appear to be promising tools through their ability to learn about the data \cite{tibshirani:1996re,friedman:2001gr,friedman:2002st,anastasiadis:2005ne} and there has been increasing interest in employing machine learning methods in medical sciences for disease classification, prognosis and prediction \cite{selaru:2002ar,seligson:2005gl,gevaert:2006pr,akay:2009su,leblanc:2010bo}.

However, the issue remains that a large number of potential stratifying variables leads to an even larger search space for treatment-variable interactions. 
Na\"{i}ve application of ML approaches fed with many irrelevant variables can lead to high dimensional models that can substantially diminish the power of a hypothesis test. 
Careful selection of potential stratifiers before discovery can therefore improve the power to detect true associations for a given effect size.

In this article, we present a framework that facilitates the incorporation of ML algorithms for treatment interaction detection in RCT data. 
We show that it is possible to use the clinical response (outcome variable) in an initial screening stage to reduce the number of stratifying variables without inflating the type-I (false positive) error rate at the interaction stage. 
In this way the interaction analysis can be applied to a targeted subspace effectively reducing the multiple comparisons in the search for interactions. 
The essence is to prioritise main effects for interaction testing over those covariates showing no evidence of association with variation in the clinical response. 
We provide theoretical results to support our assertion, making use of the randomisation of treatments in the design. 
Hence ML algorithms can informatively guide the interaction detection in order to achieve more careful expenditure of power, while ensuring that the false positive rate is controlled in a rigorous manner.

It has been proposed to use an internally developed risk model to discover TEH in RCT data \cite{burke:2014us} and it has been applied to retrospective analysis of TEH in several large RCT datasets \cite{kent:2016ri}.
This approach permits the inclusion of multiple baseline covariates to construct the internal risk model (IRM).
We provide theoretical support for this approach as it can be considered as a special case of our method.
In addition, conventional practice of the IRM approach for TEH discovery excludes the treatment variable to avoid potential biases \cite{kent:2016ri}.
This is shown to be unnecessary as a consequence of the same mathematical results that support the validity of our framework.
Furthermore, we illustrate that our framework can provide better resolution of TEH, and how that can be more sensitive to the TEH signal in certain scenarios.

\subsection{South East Asian Quinine Artesunate Malaria Trial (SEAQUAMAT)}
We demonstrate our TEH detection framework in the context of an analysis of a large anti-malaria drug trial, SEAQUAMAT \cite{dondorp:2005ar}.
The R notebooks of the interaction analyses are in the Supporting Information (SI).

SEAQUAMAT is a large randomised clinical trial that compared intravenous quinine to intravenous artesunate for treatment of severe \textit{Plasmodium falciparum} malaria.
The trial was conducted at multiple medical centers in Bangladesh, India, Indonesia, and Myanmar.
The findings of this trial provided strong evidence that intravenous artesunate had superior efficacy than intravenous quinine with one third reduction in mortality.

Severe malaria is a medical emergency with potentially a high rate of  case fatality (10 -- 40\%), therefore it is of paramount interest to understand how the treatment effect of artesunate may vary in the patient population even though its superiority of overall efficacy is well established \cite{white:2014ma}.

\section{Screening for evidence of main-effects before testing for treatment effect heterogeneity} 
\label{sec:methods}
We describe a two-stage approach to interaction discovery whereby in the  first stage variables are ranked by their main-effect strength of association with the outcome response.
This is followed by an interaction discovery stage using only the $K$ leading covariates from Stage-1.
We  show that outcome-driven screening at the first stage does not inflate the p-values obtained in the second stage testing for treatment interaction and moreover demonstrates the consistency property.

\subsection{Analysis scheme}\label{sec:analysisScheme}
For ease of exposition we restrict attention to a balanced two-arm clinical trial where individuals are randomised with equal probability onto a treatment arm $A$ versus a  standard-of-care $B$.

\subsubsection{First stage: prognostic variable screening}\label{subsubsec:pvs}

We consider an additive generalised linear regression model (GLM) for the clinical outcome,
\begin{equation}
\E{\y} = g\left(\ba + \XS\betaV + \grpT{\mathbb{I}}\theta + \XC\betaC\right),
\label{eq:main}
\end{equation}
where $\y$ denotes the clinical response for $n$ individuals recruited onto the trial, $\ba$ represents an intercept term, the $n \times p$ matrix $\XS$ contains $p$ interaction-candidate covariates so that $\betaV$ is a $p$-dimensional regression coefficient vector.  
The term $\grpT{\mathbb{I}}$ denotes an indicator binary vector of the randomised treatment allocations with elements $\grpT{\mathbb{I}}^{(i)} = 1$ if the $i$'th individual received the treatment and $\grpT{\mathbb{I}}^{(i)} = 0$ otherwise, such that $\theta$ captures the causal average treatment effect.  
The $n \times p_c$ matrix $\XC$ contains covariates, decided \textit{a priori} to be irrelevant to treatment interactions.
These could be variables that characterise the trial procedure (e.g. site of treatment) instead of the patients themselves.
The function $g\left(\cdot\right)$ is the link function that performs element-wise operation on the linear predictors.

To screen for main-effects the GLM (\ref{eq:main}) can be fit to the data to obtain a rank ordered set of variables, $\rank{X} = \left\{\rank{x_1}, ..., \rank{x_p}\right\}$, which is ordered from the most informative covariate $\rank{x_1}$ to the least informative $\rank{x_p}$.
Such ordering could be achieved through various ways, which we elaborate in section \ref{subsec:singleScreen} .

Following this, only the $K$-leading variables, $\XSK$ an $n \times K$ matrix, are selected for inclusion in the interaction discovery stage.
Note that the dimension $K$ should be decided {\em{a priori}} based on considerations of power, and not using the results of fitting the GLM.
Setting $K$ adaptively based on the evidence in $\betaVH$ from the first stage induces p-values that are not uniformly distributed under H0.
Because in that situation the degrees of freedom (DF) of the test will be subjected to random sampling.

\subsubsection{Second stage: testing for treatment interactions}

The selected covariates $\XSK$ can then be tested for treatment interactions via the following model 
\begin{equation}
\E{\y} = g\left(\alpha + {\rm{diag}}(\grpT{\mathbb{I}}) \XSK\betaAK + {\rm{diag}}(\bf{1} - \grpT{\mathbb{I}}) \XSK\betaBK  + \grpT{\mathbb{I}}\theta + \XC\betaC\right),
\label{eq:interaction}
\end{equation}
where ${\rm{diag}}(\grpT{\mathbb{I}})$ is a $n \times n$ indicator matrix with $1$'s on the diagonal elements where the corresponding vector $\grpT{\mathbb{I}}$ has a 1, and 0's elsewhere, so that we have a separate regression coefficient vector for the treatment group $\betaAK$ to that of the standard-of-care $\betaBK$.

A test for interaction can then be derived by considering the standardised magnitude of the difference vector $\left(\widehat{\betaAK} - \widehat{\betaBK}\right)$ using a test (e.g. LRT) with a corresponding p-value that is uniform under the null hypothesis ${\rm{H0}}: (\betaAK - \betaBK) = {\bf{0}}$ regardless of the fact that the variable selection in Stage-1 uses information in the outcome-driven $\betaVH$.
As a test such as the LRT provide the single p-value to indicate the evidence for interaction with treatment, it provides our framework facilitates a global test for detecting TEH.

\subsection{On the flexibility of prognostic variable screening}\label{sec:screening}

In this section we discuss various possible options for screening prognostic variables in Stage-1 (section \ref{subsubsec:pvs}).
We begin with single-stage screening procedures (section \ref{subsec:singleScreen}), followed by multi-staging screening (section \ref{subsec:pcr}), which is more generalised.
We recommend using the multi-stage screening to make the most of supervised and unsupervised ML algorithms.

\subsubsection{Single-stage screening for prognostic variables}\label{subsec:singleScreen}

We first consider the situation where variable selection is only applied once.
Possible options are:

\begin{enumerate}
\item Fitting the full additive model \eqref{eq:main}, for example in R  using {\tt{glm.R}}, and then ordering by the associated p-values assigned to the coefficients.
\item Running a LASSO model \cite{tibshirani:1996re} and then ranking by the order for which the variables enter into the LASSO path.
\item Fitting $p$ independent GLMs of the form (\ref{eq:main}) but where we include only a single variable (column) of $\XS$ at each time. 
	Then ranking covariates by the univariate p-values assigned to the columns of $\XS$ from each model.
\item Taking principle components (PCs) of $\XS$ to form a new basis design matrix $\XS^*$ with orthogonal columns made up of the PCs, and then using either option 1.\, or option 2.\, above to rank the leading PC regression coefficients.
\end{enumerate}

Option 1 is only suitable for datasets with a small or moderate number of variables, as \cite{sur:2017li} shows that in a logisitc regression, the $\chi^2$ distribution is a poor approximation of the LRT statistic when the number of variables is large in a finite dataset.
Option 2 is suitable for high-dimensional data especially when many of the covariates are not explanatory.
Option 3 is suitable for fast processing of high dimensional data, but because the selection is based on univariate regressions, it can lose power in the absence of other independent variables that can help to reduce the noise. 
Option 3 has a flavour of (univariate) filter methods for feature selection \cite{battiti:1994us,guyon:2003in}.
All of the options 1–--3 do not resolve the problem of multi-collinearity.

Application of principal component analysis (PCA) in option 4 transforms the covariates into a set of orthogonal basis, it resolves the problem of multi-collinearity. 
Hence this option is suitable for datasets having many highly correlated variables, which can occur when multiple covariates measure the same quantity.
Consequently, it enables the application of supervised methods that suffer in presence of multi-collinearity (e.g LASSO).
Although tree-based methods such as random forest can cope with multi-collinearity and therefore their performances are not affected, applying the supervised analysis on the PC scores can avoid mis-interpretation of variable importance resulted from multi-collinearity.

It is worth noting that PCA is an unsupervised approach to explain the variation in the covariates.
This means that while the most variable PC is the most informative of the shape of the covariate data cloud, it is not necessarily associated with the outcome of interest.
The advantage of our supervised approach is that it permits the variable selection to be guided by the outcome.
Thus, even if the most variable PCs are approximately noise variables, our approach is able to give priority to less variable PCs that demonstrate greater explanatory power.

There is a very interesting connection between our approach and the IRM approach \cite{burke:2014us}.
The IRM approach first constructs a regression model to explain the outcome by a set of baseline variables.
The treatment is excluded (blinded) and the model is fitted to the data from both treatment arms (instead of the control arm only). 
\begin{equation}
	\E{\y} = g\left(\ba + \XS\betaV\right),
\label{eq:irmbt}
\end{equation}
where the mathematical terms are defined the same as in equation \ref{eq:main}.
For each patient, one calculates a risk score, which defined by the linear predictor of equation \ref{eq:irmbt}.
Subsequently, another regression model is fitted containing the main effect of the treatment, the main effect of the risk scores and the effect of the interaction between the two. 
The presence of TEH is suggested by a significant effect of treatment-risk-interaction.

Like the PC scores, a risk score is also a linear projection of the baseline variables.
If $K=1$, option 4 above suggests using a supervised approach to select the most explanatory linear projections provided by PCA.
On the other hand, the IRM approach can be interpreted as selecting the most explanatory linear projection from all possible linear projections.

There are two important differences between our framework and the IRM approach.
Firstly, the main effect of treatment is included in our framework.
The IRM approach excludes the treatment assignment indicator when developing the IRM due to concerns with the potential inflation of false positive risk-treatment-interaction. 
Our mathematical argument for the validity of our framework also shows that this exclusion is unnecessary.
                                                                                                                                                                                                                                                                                                                                                                                                                                                                                                                                                                                                             
Secondly, our framework allows the strength of the interaction effect to vary across the selected variables, whereas in the risk score approach all variables composing the risks score share the same interaction effect.
Our method has the advantage of providing greater resolution of TEH. 
Consider the scenario where a (selected) variable has a relatively small main effect but has major contribution to the effect of the interaction with treatment, while the variables with dominating main effects do not interact with the treatment.
Our method can detect this signal directly because it models the individual interaction effect between the treatment and each selected variable.
However, the IRM will suffer as the dominating variables in the risk score will largely determine the shared interaction effect.
Therefore if those dominating variables do not interact with treatment, then the overall signal of treatment-risk-interaction can be diminished.

\subsubsection{Multi-stage ML screening for prognostic variables}\label{subsec:pcr}
If the ML algorithm suggests that there are $M$ variables predictive of the outcome and $M > K$, then modelling with the $K$-leading covariates may not be sufficient and potentially discarding what powerful ML methods can offer.
This can be resolved by applying dimension reduction methods such as PCA to the subset of the covariates that are found via ML to be predictive of the response.
The $K$-leading PCs will encapsulate the information from the $M$ covariates.
Stage-1 can be expanded to have multiple substages of selection:

\begin{enumerate}
	\item Substage-I: covariate selection with ML algorithm
	\begin{enumerate}
		\item An additive model is fitted to the whole dataset via a machine learning method, e.g LASSO \cite{tibshirani:1996re,zou:2005re,yuan:2006mo}, or boosting with stumps \cite{freund:1995de,friedman:2001gr}. \label{itm:mlFit}
		\item The output of step \ref{itm:mlFit} is used to rank the explanatory power of the variables, and determines $\bm{X_\mathrm{M}}$, which is an $n \times M$ matrix that contains all the variables predictive of outcome\label{itm:MLsubset}. \label{itm:MLsubset}
	\end{enumerate}	
	\item Substage-II: Selection of PCs
	\begin{enumerate}
		\item PCA is applied to $\bm{X_\mathrm{M}}$.
		\item To produce $\bm{X_{\mathrm{S}_K}}$, the $K$-leading PCs can be selected based on their score variances or the strength of association with $\y$.
	\end{enumerate}
\end{enumerate}

The positive integer $M$ can vary from sample to sample as long as $K$, is pre-defined given $n$. 
This is because the subsetting by ML in Substage-Ib of Stage-1 is independent of the interaction (to be discussed in section \ref{sec:theory} and Web Appendix A).
Therefore the DF is still kept at $K$ for the LRT in Stage-2.
Here, the scheme is described with PCA, but it can be substituted with other (unsupervised) ordination methods.

\section{Theory}\label{sec:theory}

In order to justify the proposed scheme, it requires to show that the variable selection at Stage-1 does not influence the findings of interaction detection in Stage-2. 
\begin{subgTh}
Let $\bm{\std{\hat{\beta}}}$ denote the standardised MLE of the main covariate effects in an additive GLM, while $\bm{\std{(\betaAH - \betaBH)}}$ represents the standardised MLE of the difference in treatment-specific covariate effects in an interaction model with randomised treatments $A$ and $B$.
Under a null hypothesis of ``no difference in covariate effects across treatment arms'', $\bm{\std{\hat{\beta}}}$ is independent of $\bm{\std{(\betaAH - \betaBH)}}$, such that $\bm{\std{\betaVH}} \perp \bm{\std{(\betaAH - \betaBH)}}$. 
Hence, dimension reduction via informative variable selection using $\bm{\std{\betaVH}}$ does not bias the subsequent findings of interaction discovery $\betaAH  \ne \betaBH$.
\label{theorem:subg}
\end{subgTh}
Proof: The proof is contained in Web Appendix A.
The proof also shows that theorem 1 holds after a linear transformation on the covariates.



In practice, the requirement of prognostic screening at Stage-1 depends on the power we have in the test for interaction in Stage-2.
If there is sufficient power to perform a global test with all $p$ variables then Stage-1 can be bypassed and therefore $K = p$.

If the power at Stage-2 is insufficient to include all $p$ variables, then prognostic screening is performed to select $K$ variables. 
$K$ should be a deterministic function of $n$ that is monotonic increasing with no upper bound, such as $\log(n)$ or power curve based on $n$ (but independent of the sampled covariate values $\bm{X}$ as they vary from sample to sample). 
Permitting $K$ to increase with $n$ ensures the consistency of the procedure. 
That is even with the worst case scenario where true interaction occurs at the least prognostic variable $\bm{x_j}$, as $n$ tends to infinity, $K$ will eventually reach $p$ to include $\bm{x_j}$. 
The property of such a scheme relies on the test in Stage-2 to be consistent, such as the LRT. 
Theoretically, this scheme can even withstand the cryptic interactions that leads from $\beta_{Aj} \ne \beta_{Bj}$ but $\beta_j = 0$.

In section \ref{subsec:pcr} we have discussed a screening procedure with two substages.
In Substage-I, a chosen supervised approach selects $M$ covariates, $\bm{X_\mathrm{M}}$, based on the evidence for their prognostic effect.
And subsequently in Substage-II, we select the $K$-leading PCs of $\bm{X_\mathrm{M}}$ from Substage-I.

If the selection of the PCs is based on the ranking of PC score variance, then it is independent of outcome. 
Hence the outcome-driven dimension reduction is only executed in Substage-I to obtain $\bm{\beta_\mathrm{M}}$, and theorem 1 states this selection does not influence the subsequent interaction test. 

If the PC selection is based on the ranking of their strength of association with the outcome, then the outcome-driven dimension reduction is performed in both substages.
The covariates $\bm{X_\mathrm{M}}$ are selected in Substage-I as above.
PCs of $\bm{X_\mathrm{M}}$ is a linear transformation.
Since theorem 1 holds after a linear transformation on the covariates, this means that the PCs of $\bm{X_\mathrm{M}}$ can be selected via a supervised approach to obtain $\XS$ in substage-II without biasing the test in Stage-2 that detects the interaction between treatment and $\XS$.


Lets consider all possible linear transformations.
Because theorem 1 is stable under linear transformation, if we select the most prognostic linear combination of the covariates, it will not induce bias the subsequent interaction test. 
So a consequence of our proof is that if the treatment is included in a GLM risk model as in equation \ref{eq:main}, it will not inflate the false positive rate of risk-treatment-interaction.
In that case the GLM risk model would take the form in equation \eqref{eq:main}, and the risk score would be defined as $\XS\betaV$.

In fact, given a set of variables, one can perform any transformation or basis expansion to generate a new set of bases as the interaction candidates.
As long as these operations do not depend on a function of the treatment indicator, the new set of basis can be fed into our framework described in section \ref{sec:analysisScheme} without inflating the false positive rate.
This enables the accommodation of non-linearity, and moreover allows the employment of various powerful ML methods such as boosting \cite{friedman:2001gr}.

\section{Empirical study}
We demonstrate our proposed scheme for interaction detection on the RCT dataset from SEAQUAMAT \cite{dondorp:2005ar}.
Detailed description of the variables and the following analyses are in the R notebooks in SI.

We first describe the data preparation required for the analysis (section \ref{sec:processData}).
Subsequently we present the analysis including all the biological variables in the interaction test, and illustrate its lack of power, which motivates the development of our methodology.
As mentioned in section \ref{sec:screening}, we prefer the multi-stage screening procedure and therefore this screening procedure is chosen for our primary analyses presented in section \ref{sec:primaryAnalysis}.
There we demonstrate that power can be gained from adopting our framework for TEH discovery.

The full details of the analyses in this section can be found in R notebook S1: Empirical demonstration of interaction analysis on full malaria trial dataset.

\subsection{Pre-process of data}\label{sec:processData}
The dataset is first examined to determine any processing required prior to the analysis.
Missing values are imputed by the R package MICE \cite{buuren:2011mi} and 50 imputed datasets are created.
The analysis is performed on each imputed dataset to investigate whether the findings are consistent across imputation variation.

\subsection{Interaction analysis with GLM including all variables}
In this dataset there are 14 biological variables that serve as interaction-candidates.
While it is understandable that treatment site (country) may have an overall effect, it is not so intuitive to expect it to interact with treatment.
Therefore, country is included in the model for overall adjustments, but not included as an interaction-candidate.

We fit the full additive model $\mathcal{M}^\mathrm{(F)}_\mathrm{{A}}$ and the full interaction model $\mathcal{M}^\mathrm{(F)}_\mathrm{{I}}$. 
So that $\mathcal{M}^\mathrm{(F)}_\mathrm{{A}}$ includes the main effect of all variables,
while $\mathcal{M}^\mathrm{(F)}_\mathrm{{I}}$ also includes all the interaction terms between treatment and the 14 biological variables.
For each imputed dataset, an LRT is performed to test whether $\mathcal{M}^\mathrm{(F)}_\mathrm{{A}}$ is an adequate approximation of $\mathcal{M}^\mathrm{(F)}_\mathrm{{I}}$.
The median LRT p-value is 0.115 (p-value range = [0.0775, 0.159], 3 s.f.\footnote{$x$ s.f. is short for rounding to $x$ significant figures.}), only 9 of the 50 imputed datasets have ANOVA p-values $<0.1$.
It is apparent that there is no evidence of treatment interaction in SEAQUAMAT dataset, which may be due to lack of power for testing all 14 variables simultaneously.

\subsection{Interaction discovery with PC-GLM after outcome-driven dimension reduction}\label{sec:primaryAnalysis}
After data preparation (section \ref{sec:processData}), we do not observe high correlations among the covariates. 
(Further details are provided in the data preparation section in the R notebook S1.) 
Therefore, it is more appropriate to perform the supervised variable selection on the covariates directly rather than the PCs.
This supports our choice of multi-stage screening.

Recall that multi-stage screening divides Stage-1 into substages.
Section \ref{sec:pavs} describes Substage-I, where we perform supervised variable selection to subset the variables.
Then in Substage-II, PCA is applied to this subset of variables.
We present two variants of this analysis, where the $K$-leading PCs are selected based on (i) their variances (section \ref{subsubsubsec:uspca}) and (ii) supervised selection guided by outcome (section \ref{sec:spca}).

\subsubsection{Supervised screening by the gradient boosting model}\label{sec:pavs}
To screen the covariates, an additive model is fitted by a gradient boosting model (GBM) \cite{friedman:2001gr,friedman:2002st} with stumps by using the R package gbm \cite{ridgeway:2017gb}.
The covariates are ranked according to the relative influence (RI) values.

Of the 50 imputed datasets, 37 supports a model with treatment, country and additional 10 biological covariates.
These biological covariates are consistent cross the 37 imputed datasets and these are coma, logged parasitaemia, age, heart rate, temperature ($^\circ\text{C}$), systolic blood pressure, haematocrit, weight, jaundice, duration of fever (days).

\subsubsection{Unsupervised PC selection}\label{subsubsubsec:uspca}
From the additive model fitted using GBM with stumps in section \ref{sec:pavs}, we have identified a subset of 10 biological variables that have the greatest explanatory power in the majority (37/50) of the imputed datasets.
PCA is subsequently performed on this subset and the PCs are ranked by their PC score variances.
Here, we test whether there are differences in PC effects between treatment groups. 
Again, for the purpose of demonstrating our method, we iterate through values of $K$ from 1 to 10 for the LRT in the Stage-2. 

\begin{figure}[ht]
  \centering
    \includegraphics[width=0.8\textwidth]{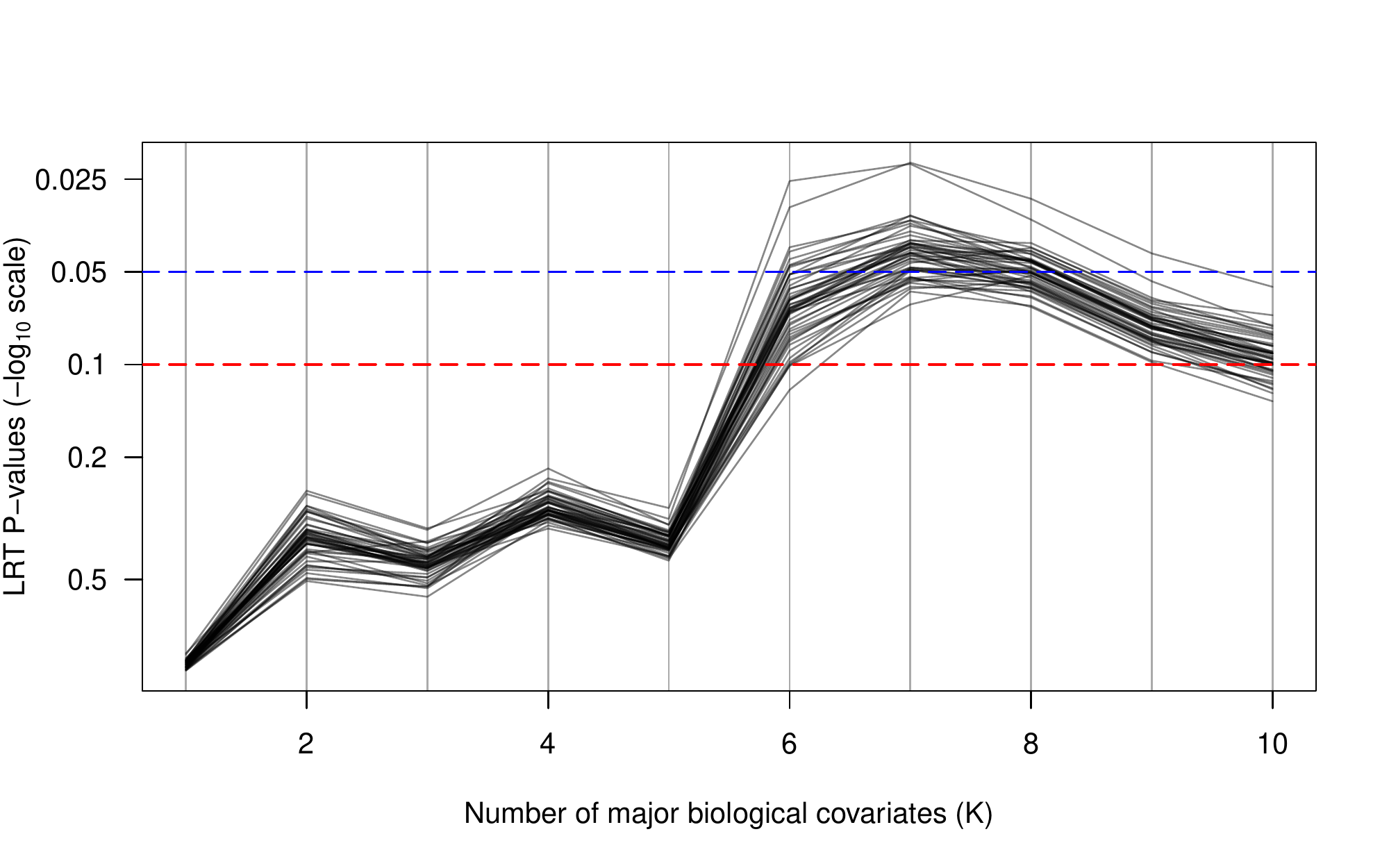}
  \caption{The LRT p-values versus degrees of freedom ($K$) across 50 imputed datasets of all 1461 patients recruited onto the SEAQUAMAT trial. 
  Here, $K$ represents the number of most variables PCs included in test for interaction with treatment.}
  \label{fig:malariaPCARegDFs}
\end{figure}

The results from testing the interactions between treatment and the leading PCs exhibit a reasonably stable pattern as portrayed in figure \ref{fig:malariaPCARegDFs}. 
There is generally no evidence of interactions between treatment and the $K$-leading PCs for $\mathrm{DF} = K = {1, ..., 5}$, where all of the imputed datasets have p-value $>$ 0.1.
The number of imputed datasets with p-value $<$ 0.1 is respectively 47, 50, 50, 50 and 34 for including the leading 6, 7, ..., 10 PCs.
The overall evidence for interactions is strongest when testing for interaction between treatment and the 7 leading PCs. 
The evidence gradually weakens as more PCs are added, suggesting the minor PCs are not so informative of TEH.

\begin{figure}[ht]
  \centering
    \includegraphics[width=0.7\textwidth]{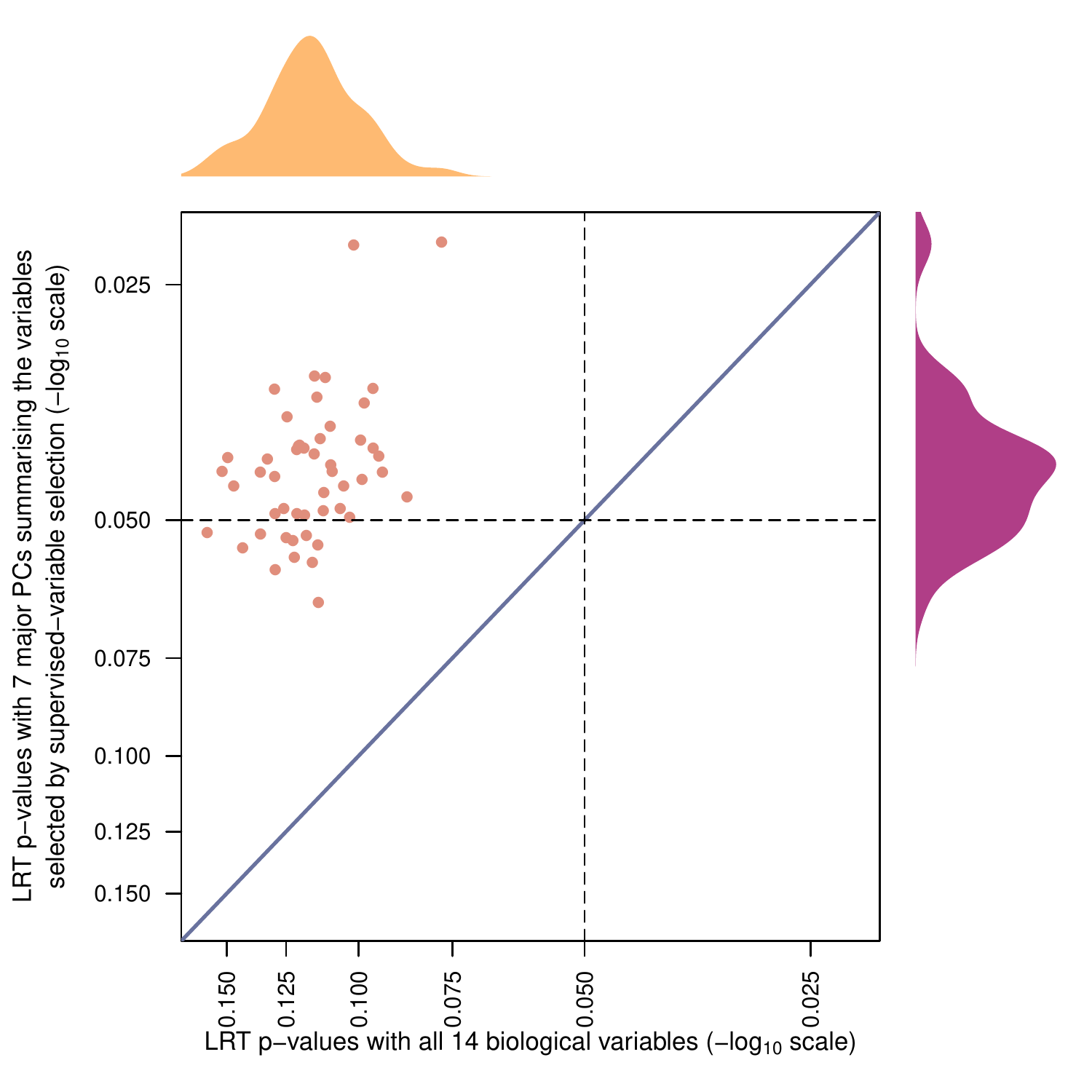}
  \caption{Comparison of LRT P-values in a PC-GLM framework for the 1461-patient SEAQUAMAT data. 
  For every of the 50 imputed dataset, the LRT p-values for TEH are compared between the analysis with all 14 biological variables and that with the 7 leading PCs that represent all the variables picked by supervised-variable selection (GBM) in Substage-I of Stage 1.
  The purple straight line goes through the origin and has a slope of 1.}
  \label{fig:malariaPCAGLMLRTPvals}
\end{figure}

When including the 7 leading PCs in the interaction test, where the median p-value = 0.0435 (p-value range = [0.0221, 0.0638]) and p-value $<$ 0.05 for 39 of the 50 imputed dataset.
Moreover, figure \ref{fig:malariaPCAGLMLRTPvals} illustrates that the evidence for TEH is noticeably stronger (LRT p-value decreases) for every imputed dataset when we reduce the variables tested from all 14 biological variables to the 7 PCs representing the 10 variables selected by GBM (in section \ref{sec:pavs}).

There is a notable increase in the strength of evidence for interactions when PC6 is included to test for interaction with treatment. 
From inspecting the correlations between PC6 and biological variables, we find that PC6 is most strongly correlated with jaundice and then systolic blood pressure.
Furthermore, jaundice and systolic blood pressure have the largest absolute PC6 coefficients across all imputed datasets.
The signal for interactions attributed to jaundice and systolic blood pressure appears to be captured by PC6.

The variables identified here, jaundice and then systolic blood pressure, that seems to be responsible for TEH, are consistent with the results from the analysis with single-stage screening (B.1.1 of Web Appendix B), where the evidence for interactions strengthens when jaundice and systolic blood pressure are added in turn.
Comparing figure \ref{fig:malariaPCAGLMLRTPvals} with Web Figure B.2 (in Web Appendix B), the power to detect interactions appears to greater when using multi-stage screening than single-stage screening.

In summary, this analysis clearly demonstrates the how power can be gained after supervised selection of variables based on the evidence for the main effects without segregation by treatment arms.

\subsubsection{Supervised PC selection} \label{sec:spca}
For each imputed dataset, an additive LASSO model is constructed to explain the outcome by the treatment and the PCs obtained in PCA (described in \ref{subsubsubsec:uspca}).
This provides a ranking of the PCs by the explanatory power of their main effects.
Therefore, $K$ represents the most explanatory PCs informed by LASSO.
To explore the property of this analysis design, we iterate through values of $K$ from 1 to 10 when performing the interaction test in Stage-II.

\begin{figure}[ht]
  \centering
    \includegraphics[width=0.8\textwidth]{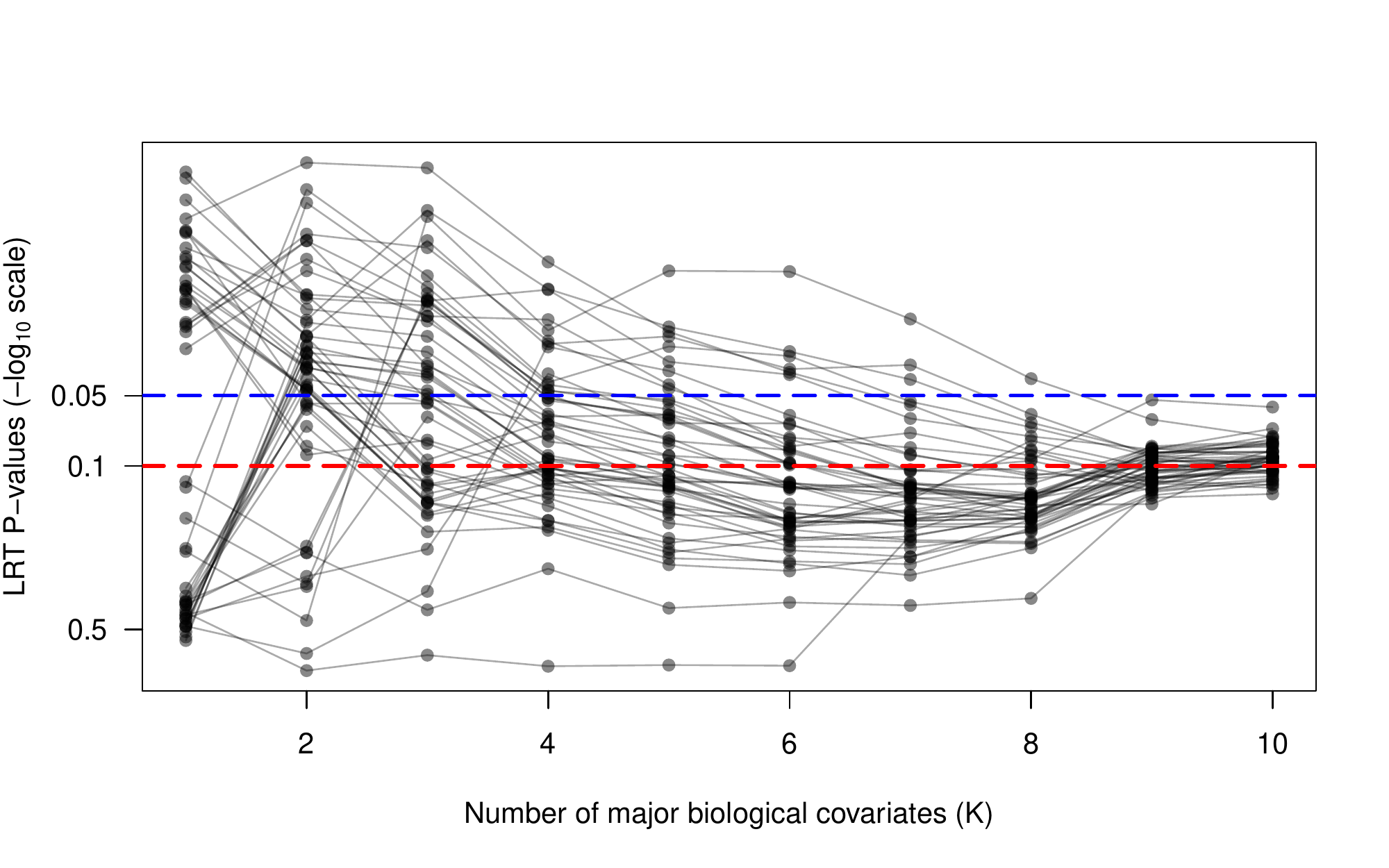}
  \caption{The LRT p-values versus degrees of freedom ($K$) across 50 imputed datasets of all 1461 patients recruited onto the SEAQUAMAT trial. 
  Here, $K$ represents the number of most variables PCs included in test for interaction with treatment.}
  \label{fig:malariaSupPCARegDFs}
\end{figure}

Figure \ref{fig:malariaSupPCARegDFs} presents the LRT p-values obtained from each imputed datasets for $K = \left\{1, ..., 10\right\}$.
In general, it appears that the power is the strongest when we include the $K = 2$ most explanatory PCs in the interaction test.
In this case, 34/50 of the imputed datasets have LRT p-values $<$0.05, and 41/50 of those have LRT p-values $<$0.1.
For these 41 imputed datasets, PC6 is found in the $K = 2$ PCs selected for the interaction test.
This indicates that PC6 is most informative of TEH, which is consistent with the analysis presented in sections \ref{subsubsubsec:uspca}, which indicates that PC6 is most informative of TEH.

The pattern of p-values across $K$ is much more noisy in figure \ref{fig:malariaSupPCARegDFs} than that in figure \ref{fig:malariaPCARegDFs}.
This is due to the variation in the LASSO ranking of PCs across imputed datasets.
Once we have included the most two explanatory PCs in the interaction test, inclusion of additional PCs generally leads to drop of power.
This is because, in this case PC6 is most informative of TEH and it is also (nearly) most explanatory.
However, it is important to note that this phenomenon is not necessarily the case as we have already explained in section \ref{sec:theory} and the proof in Web Appendix A.

\subsubsection{Supplementary analyses}
Aside from the primary analysis presented in section \ref{sec:primaryAnalysis}, we have also performed additional analyses on the same dataset to further explore our method.
These supplementary analyses are presented in Web Appendix B, containing two sections: B.1 and B.2.
In section B.1, we present a variant of the analysis in \ref{sec:primaryAnalysis}, where we employ single-stage screening instead of the preferred multi-stage screening.
In section B.2, we present the analyses on subsets of the SEAQUAMAT trial data to investigate the behaviour of our method when the sample size is substantially reduced.

\section{Discussion}

In this paper we present a framework that increases the power of interaction discovery by outcome-driven selection of variables without inflating the type I error rate.
Using the severe malaria trial data as a motivating study for our method, there appears to be TEH of artesuate. 
While the interaction test we perform is global, we are still able to decompose the signal, which suggests that the TEH may be attributed to jaundice and systolic blood pressure.

Interaction discovery plays a key role in stratified medicine and is a challenging task as soon as the number of interaction candidate exceeds a handful of variables.
The limited power for interaction discovery forbids a carpet search for the interaction effects.
This conflicts with the desire to learn as much as possible from the data and stresses the need for global hypothesis testing tools for interaction discovery.
Our novel screening procedure is fairly flexible and hence can take advantage of powerful ML algorithms.
While we have focused on application to clinical trial data in this paper, our flexible framework can be applied to datasets from any equivalent randomised experiments for retrospective exploration of TEH.

ML algorithms themselves lack the statistical rigour required for hypothesis testing.
In spite of this, our mathematical results support the validity of how ML algorithms are incorporated in our hypothesis testing framework.

In line with the recommendations on interaction analyses that demands hypotheses to be pre-defined \cite{rothwell:2005su,kent:2010as,sun:2014us,burke:2015th}, our framework requires determining \textit{a priori} the degree of freedom ($K$) of the test for interaction effect in Stage-2.
This is an important reminder that there is no free power.
The decision on the degrees of freedom and the rest of the analysis would be recorded in a scientific notebook such as an R notebook \cite{allaire:2017rm}.

In contrast to the strictly hypothesis driven interaction analyses, our framework relaxes some of the stringencies as it does not require explicit specification of the (few) factors to be tested.
Instead we can initiate with a much larger pool of candidates, and then narrow it down by variable selection and dimension reduction by ML algorithms.
Furthermore, the highly flexible ML algorithms have much weaker assumptions on the properties of the underlying random process than traditional methodologies.
Therefore it allows the data to speak for itself.

We found that our approach can be considered as a generalisation of the internal risk model approach.
Because our framework facilitates TEH discovery with finer resolution by allowing the interactions between the treatment and the variables to be modelled individually.
Therefore it can better detect TEH when the variables driving TEH have weaker signal for main effect than dominating prognostic variables that do not contribute to TEH.

The statistical evidence for TEH provided by this method is at the global level and not for the variables individually.
If global signal is decomposed to pin down the potential set of variables responsible for treatment interaction, it should be treated with caution, specifically transformation has been applied.

In our demonstrations, PCA was employed for dimension reduction, which performs a linear transformation.
By examining the PC coefficients or the correlations between the PCs scores and variables of interest, it allows us to hypothesise which variables interact with the treatment.
However, if there are concerns regarding the interpretability of an analysis with a transformation, then the user need to design the analysis based on what is more important.

The user should always use the transformation that is most meaningful to the problem of interest.
If the purpose of the analysis is to purely indicate whether or not there is evidence for TEH, then the user only need to interpret the LRT p-value, so it might be more appropriate to choose a transformation that would achieve the most thorough exploration of the data.
However, if the aim of the analysis is to generate hypotheses on which specific variables contribute to TEH, then they should employ the transformation (including no transformation) that would be most helpful for deciding which variables to be tested in future experiments.
Of course, there is a trade-off---a simpler analysis would be easier to interpret but it can potentially miss the complex patterns and hence the signal in the data.
This is one of the examples of ``no-free-lunch'' in statistical analysis.
It is also important to note that if the user decomposes the global/aggregated interaction signal to obtain a suggested set of original variables driving the interaction with treatment, the identification of an interaction between the treatment and a specific variable should be viewed as hypothesis generation. 

Our framework furnishes a more holistic exploration of the data while also provides rigorous evidence for potential interactions in retrospective interaction analyses.
Although the false positive rate is controlled in our approach, the findings from using this framework will still need to be replicated in other datasets for validation \cite{rothwell:2005su}.
However, our method can assist researchers to decide whether to further invest in exploring the heterogeneous effect of a treatment.

\section*{Acknowledgements}
C.-H. Wu is funded by Medical Research Council UK and Cancer-Research UK (Grant no.: ANR0031).
We thank Mahidol University Oxford Tropical Medicine Research Unit (MORU) for providing us with the data from the SEAQUAMAT trial. C. C. Holmes is supported by the Medical Research Council UK, the EPSRC, the Alan Turing Institute, and the Li Ka Shing Centre for Health Innovation and Discovery.

\bibliographystyle{unsrt} 
\bibliography{variableSelectTEH}

\newpage

\appendix
\renewcommand{\figurename}{Web Figure}
\renewcommand\thefigure{\thesection.\arabic{figure}}

\section*{Supplementary Materials}

\section{Web Appendix A: Mathematical arguments for theorem 1}\label{appA:proof}
\setcounter{figure}{0}   

In this section, we present the proof that under a null hypothesis, H0, of  ``no difference in covariate effects across treatment arms'', in a generalised linear model (GLM) framework, the standardised MLEs of the coefficients $\bm{\std{\betaVH}}$ in an additive model are (asymptotically) independent of the standardised MLE difference between the group-specific coefficients,  $\std{(\bm{\betaAH - \betaBH})}$ in an treatment interaction model, when we have RCT data.

\subsection*{Defining the models and parameters}
Let $n$ be the number of individuals, where $n = n_\mathrm{A} + n_\mathrm{B}$, with $n_g$ indicating the number individuals in treatment group $g \in \left\{\mathrm{A}, \mathrm{B}\right\}$. 
The $n \times p$ matrix $\XS$ contains the $p$ interaction-candidate covariates, that potentially contribute to the interaction effect.
In a study, we might also observe $p_c$ covariates, which we do not consider as interaction-candidates.
The $n \times p_c$ matrix $\XC$ holds the $p_c$ interaction-irrelevant covariates, whose coefficient values are ``c''ommon across treatment groups.
The term $\grpA{1}$ and $\grpB{1}$ are vectors of length $n_\mathrm{A}$ and $n_\mathrm{B}$ respectively, where all elements equal to 1.
On the other hand, $\grpA{0}$ and $\grpB{0}$ are the equivalent 0 vectors.
The outcome values of all $n$ patients are recorded in the vector $\y$, and are assumed to have come from a distribution in the exponential family.

For the purpose of the following calculations, the outcome vector and covariate matrices are arranged such that 
\begin{equation}
\bm{y} = \left(\!
    \begin{array}{c}
      \yA \\
      \yB
    \end{array}
  \!\right), \quad
\bm{X_\mathrm{S}}= \left(\!
    \begin{array}{c}
      \XA \\
      \XB
    \end{array}
  \!\right), \quad
\bm{\XC} = \left(\!
    \begin{array}{c}
      \XCA \\
      \XCB
    \end{array}
  \!\right), 
\label{eq:dataByGroup}
\end{equation}
where $\bm{y_g}$, $\bm{X_g}$ and $\bm{X_{\mathrm{C}g}}$ are respectively the outcome vector and covariate matrices of treatment group $g$. 

Consider a GLM of the following form
\begin{equation}
\mathbb{E}[y_i]	= g(\eta_i), \nonumber
\end{equation}
where $y_i$ denotes the $\ith{i}$ element of $\y$, $g(\cdot)$ is the link function and $\eta_i$ is the linear predictor of the $\ith{i}$ observation. The vector $\bm{\eta}$ represents the linear predictor for all observations. 

\subsubsection*{Additive model}

To set up the additive model (no interaction), $\mathcal{M}_\mathrm{A}$ we first construct the data matrix $\bm{D}$ as
\begin{equation}
\bm{D} = \left(\!
    \begin{array}{cccc}
      \bm{\grpA{X}} &  \grpA{1} & \grpA{0} & \XCA\\
      \bm{\grpB{X}} &  \grpB{0} & \grpB{1} & \XCB
    \end{array}
  \!\right).
\label{eq:matD}
\end{equation}
This column arrangement has been selected because our primary interest is in the interaction-candidate covariates.
The linear predictor of the additive GLM is defined as
\begin{align}
	\bm{\eta} &= \bm{X}\bm{\gamma}, \nonumber \\
	\X &= \bm{D}, 
\end{align}
where the coefficient vector $\bm{\gamma}$ can be decomposed into $\bm{\gamma} = \left(\betaV, \beta_\mathrm{A0}, \beta_\mathrm{B0}, \betaC \right)$. 
The $\betaV$ is a vector the $p$ pooled main effects of the interaction-candidates, while $\betaC$ is the corresponding vector for the $p_c$ interaction-irrelevant covariates.
The term $\beta_{g0}$ is the intercept of treatment group $g \in \left\{\mathrm{A}, \mathrm{B}\right\}$.
The matrix $\X$ appears redundant here, however, having this `wrapper' matrix enables more flexible construction of the linear model as we demonstrate in a later section on PC-GLM.

\subsubsection*{Interaction model}

To construct the interaction model, $\mathcal{M}_{I(1)}$, the matrix $\bm{D}$ is expanded subsequently to obtain
\begin{equation}
\one{D} = \left(\!
    \begin{array}{ccccc}
      \bm{\grpA{X}} &  \bm{0_{n_\mathrm{A} \times p}} & \grpA{1} & \grpA{0} & \XCA\\
      \bm{0_{n_\mathrm{B} \times p}} & \bm{\grpB{X}} &  \grpB{0} & \grpB{1} & \XCB
    \end{array}
  \!\right), 	
\label{eq:matD1}
\end{equation}
where $\one{D}$ is a $n \times (2 (p+1) + p_c)$ matrix and $\bm{0_{l_1 \times l_2}}$ is a $l_1 \times l_2$ zero matrix.

The linear predictor of a full interaction model with treatment can be constructed as
\begin{align}
\bm{\eta} &= \one{X}\one{\gamma},\nonumber\\
\one{X} &= \one{D}.
\label{eq:intModel1}
\end{align}
The coefficient vector $\one{\gamma}$ is defined as $\one{\gamma} = \left(\betaA, \betaB, \tilde{\beta}_\mathrm{A0}, \tilde{\beta}_\mathrm{B0}, \bm{\tilde{\beta}_C} \right)$.
The vector $\bm{\beta_g}$ is the treatment-specifc  coefficients of the interaction-candidates for treatment group $g \in \left\{\mathrm{A}, \mathrm{B}\right\}$.
The terms $\tilde{\beta}_\mathrm{A0}$, $\tilde{\beta}_\mathrm{B0}$ and $\bm{\tilde{\beta}_C}$ are equivalent to $\beta_\mathrm{A0}$, $\beta_\mathrm{B0}$ and $\betaC$ respectively.

In preparation of an alternative construction, the columns of $\one{D}$ are re-arranged to obtain: 
\begin{equation}
\two{X} = \left(\!
   \begin{array}{ccccc}
      \bm{0_{n_\mathrm{A} \times p}} & \bm{\grpA{X}} &  \grpA{0} & \grpA{1} & \XCA\\
      \bm{\grpB{X}} &  \bm{0_{n_\mathrm{B} \times p}} & \grpB{1} & \grpB{0} & \XCB
    \end{array}
  \!\right). 	
\label{eq:matD2}	
\end{equation}
With this matrix, the linear predictor of an alternative interaction model, $\mathcal{M}_{I(2)}$, is set up according to 
\begin{align}
\bm{\eta} &= \two{X}\two{\gamma},\nonumber\\
\two{X} &= \two{D}.
\label{eq:intModel2}
\end{align}
where the coefficient vector $\two{\gamma} = \left(\betaB, \betaA, \tilde{\beta}_\mathrm{B0}, \tilde{\beta}_\mathrm{A0}, \bm{\tilde{\beta}_C} \right)$.
Note that $\mathcal{M}_{I(1)}$ and $\mathcal{M}_{I(2)}$ have the same column space, and hence their respective coefficient vectors are composed of the same set of elements.

In this case, it is observed that the matrix $\X$ can be defined in terms of linear projections of $\one{X}$ and $\two{X}$, according to
\begin{align}
\X &= \one{X} \one{J} \nonumber\\
   &= \two{X} \two{J}, 
\label{eq:Js}
\end{align}
with
\begin{equation}
\one{J} = \left(\!
    \begin{array}{cccc}
      \bm{\mathrm{I}_p} & \multicolumn{3}{c}{\bm{0_{p \times (p_c + 2)}}}\\
      \bm{\mathrm{I}_p} & \multicolumn{3}{c}{\bm{0_{p \times (p_c + 2)}}}\\
      \bm{\transpose{0_p}} & 1 & 0 & \bm{\transpose{0_{p_c}}}\\
      \bm{\transpose{0_p}} & 0 & 1 & \bm{\transpose{0_{p_c}}}\\
      \multicolumn{3}{c}{\bm{0_{p_c \times (p+2)}}} & \bm{\mathrm{I}_{p_c}}
    \end{array}
  \!\right), \quad
\two{J} = \left(\!
    \begin{array}{cccc}
      \bm{\mathrm{I}_p} & \multicolumn{3}{c}{\bm{0_{p \times p_c + 2}}}\\
      \bm{\mathrm{I}_p} & \multicolumn{3}{c}{\bm{0_{p \times p_c + 2}}}\\
      \bm{\transpose{0_p}} & 0 & 1 & \bm{\transpose{0_{p_c}}}\\
      \bm{\transpose{0_p}} & 1 & 0 & \bm{\transpose{0_{p_c}}}\\
      \multicolumn{3}{c}{\bm{0_{p_c \times (p+2)}}} & \bm{\mathrm{I}_{p_c}}
    \end{array}
  \!\right), 
\nonumber
\end{equation}
where both $\one{J}$ and $\two{J}$ are $(2(p + 1) + p_c) \times (p + p_c +2)$.
The notation $\bm{\mathrm{I}_d}$ represents a $d \times d$ identity matrix.
The expression $\bm{0}_d$ is a $d$-dimensional zero vector, while $\bm{0}_{d_1 \times d_2}$ is a $d_1 \times d_2$ zero matrix.

\subsection*{Calculations}
We first demonstrate $\betaVH \perp \betaAH - \betaBH$, which can be used to argue that $\bm{\std{\betaVH}} \perp \bm{\std{(\betaAH - \betaBH)}}$ as $n \rightarrow \infty$.
The MLEs of $\gammaV$, $\one{\gamma}$ and $\two{\gamma}$ are respectively $\gammaVH$, $\one{\gammaVH}$ and $\two{\gammaVH}$. 
Asymptotically, $\gammaVH$ follows a multivariate normal distribution, $\gammaV \sim \mathcal{N}_{p+p_c+2}(\gammaV, \bm{\Sigma})$, with the variance-covariance matrix $\bm{\Sigma}$.
Similarly, $\one{\gammaVH} \sim \mathcal{N}_{p'}(\one{\gamma}, \one{\Sigma})$, with $p' = 2p+p_c+2$.
As $\one{\gammaVH} - \two{\gammaVH}$ is a linear transformation of $\one{\gammaVH}$, the difference vector also follows a normal distribution, albeit only having $2p$ dimensions.
This is due to the zeros in the last $p_c + 2$ elements of $\one{\gammaVH} - \two{\gammaVH}$.

We use the notation $\bm{M}_{r_1:r_2 \times c_1:c_2}$ to refer to the sub-matrix from rows $r_1$ to $r_2$ and from columns $c_1$ to $c_2$ of the original matrix $\bm{M}$.
The covariance matrix $\Cov{\betaVH}{\betaAH - \betaBH}$ is located in the sub-matrix $\Cov{\gammaVH}{\one{\gammaVH} - \two{\gammaVH}}_{1:p \times 1:p}$.
By the properties of normal random variables, $\text{Cov}(\gammaVH, \one{\gammaVH} - \two{\gammaVH})_{1:p \times 1:p} = \bm{0_{p \times p}}$ implies $\betaVH$ and $ \betaAH - \betaBH$ are independent.

The MLEs of GLM coefficients are often computed using the iteratively re-weighted least squares (IRLS) algorithm.
The estimates of $\gammaV$, $\one{\gamma}$ and $\two{\gamma}$ respectively take the form
\begin{align}
\bm{\hat{\gamma}} &= \left(\XT\bm{\hat{W}}\X\right)^{-1}\XT\bm{\hat{W}z}, \nonumber \\
\bm{\hat{\gamma}_{(1)}} &= \left(\XT\bm{\hat{W}_{(1)}}\X\right)^{-1}\XT\bm{\hat{W}_{(1)} z_{(1)}}, \nonumber \\
\bm{\hat{\gamma}_{(2)}} &= \left(\XT\bm{\hat{W}_{(2)}}\X\right)^{-1}\XT\bm{\hat{W}_{(2)} z_{(2)}}, \nonumber
\end{align}
$\bm{\hat{W}}$ is the MLE estimate of $\bm{W}$, which is an $n \times n$ diagonal matrix, and the $\ith{i}$ diagonal element, $w_i$, is the weight of the $\ith{i}$ individual given the data and model parameters.
$\bm{\hat{z}}$ denotes the MLE the $n$-dimensional vector $\bm{z}$, which is the working dependent variable given the data and model parameters.
In the interaction models, $\model{I(1)}$ and $\model{I(2)}$, the terms $\one{\hat{W}}$, $\two{\hat{W}}$, $\one{\hat{z}}$ and $\two{\hat{z}}$ are defined similarly.

The covariate matrix $\Cov{\gammaVH}{\one{\gammaVH} - \two{\gammaVH}}$ can be expressed as
\begin{align}
& \mathrm{Cov} \text{\Large\{\normalsize} \left(\XT \WH \X\right)^{-1}\XT \WH\bm{z}, \nonumber \\ 
& \quad  \left(\oneT{X} \one{\WH} \one{\X}\right)^{-1}\oneT{X}\one{\WH}\one{z} - \left(\twoT{X} \two{\WH} \two{\X}\right)^{-1}\twoT{X}\two{\WH}\two{z} \text{\Large\}\normalsize}
\label{eq:covEx1}
\end{align}

Since $\one{\X}$ and $\two{\X}$ have the exact same set of column vectors, we have $\one{\hat{W}} = \two{\hat{W}} = \bm{\widetilde{W}}$ and $\one{z} = \two{z} = \bm{\widetilde{z}}$.
We can then rewrite equation \eqref{eq:covEx1} as 
\begin{align}
& \left(\XT \WH \X\right)^{-1}\XT\WH \Cov{\z}{\zt} \Wt \one{X} \left(\oneT{X} \Wt \one{\X}\right)^{-1} -  \nonumber\\
& \quad \quad \quad \quad \quad \left(\XT \WH \X\right)^{-1}\XT\WH \Cov{\z}{\zt} \Wt \two{X} \left(\twoT{X} \Wt \two{\X}\right)^{-1}\nonumber\\
=& \bm{Q}\left[{\Wt \one{X} \left(\oneT{X} \Wt \one{\X}\right)^{-1} - 
\Wt \two{X} \left(\twoT{X} \Wt \two{\X}\right)^{-1}}\right]\nonumber\\
=& \bm{Q}\left(\X\XT\right)^{-1}\X\left[\XT{\Wt \one{X} \left(\oneT{X} \Wt \one{\X}\right)^{-1} - 
\XT\Wt \two{X} \left(\twoT{X} \Wt \two{\X}\right)^{-1}}\right],
\label{eq:covEx2}
\end{align}
where $\bm{Q} = \left(\XT \WH \X\right)^{-1}\XT\WH \Cov{\z}{\zt}$.
%
%
This can be further simplified by substituting equation \eqref{eq:Js} into equation \eqref{eq:covEx2}, leading to 
\begin{equation}
\bm{Q}\left(\X\XT\right)^{-1}\X\left(\bm{\transpose{J}_{(1)}} - \bm{\transpose{J}_{(2)}}\right).
\label{eq:covEx4}
\end{equation}
Because the first $p$ columns of $\one{\transpose{J}} - \two{\transpose{J}}$ are zero columns, so are those in the matrix $\Cov{\gammaVH}{\one{\gammaVH} - \two{\gammaVH}}$ and hence $\Cov{\betaVH}{\betaAH - \betaBH}$ is a zero matrix.

When the underlying model is correct, as $n \rightarrow \infty$, the MLE of a parameter approaches a normal distribution and its standard error approaches the asymptotic standard deviation of the estimator.
Consequently, a standardised MLE will be asymptotically normally distributed.
Therefore, the zero-covariance property is again a sufficient condition for independence.
Let $\bm{\std{\gammaVH}}$ and $\bm{\std{(\one{\gammaVH} - \two{\gammaVH})}}$ denote the standardised MLEs of $\bm{\gammaVH}$ and $\bm{\one{\gammaVH} - \two{\gammaVH}}$ respectively.

The covariance matrix of the standardised MLEs is 
\begin{equation}
\text{Cov}\left(\bm{\std{\gammaVH}}, \bm{\std{(\one{\gammaVH} - \two{\gammaVH})}}\right) = \text{diag}(\bm{\text{SE}_{\gammaVH}})^{-1}\text{Cov}\left(\gammaVH, \one{\gammaVH} - \two{\gammaVH}\right)\text{diag}(\bm{\text{SE}_{\one{\gammaVH} - \two{\gammaVH}}})^{-1},\nonumber
\end{equation}
where diag($\bm{v}$) is a diagonal matrix with $\ith{i}$ diagonal value equal to the $\ith{i}$ element of some vector $\bm{v}$.
The $\ith{i}$ element of the vector $\mathrm{SE}_{\gammaVH}$ is the standard error of the $\ith{i}$ element of $\gammaVH$.
The standard error vector $\bm{\text{SE}_{\one{\gammaVH} - \two{\gammaVH}}}$ for $\one{\gammaVH} - \two{\gammaVH}$ is similarly defined.

As the first $p$ columns of $\text{Cov}(\gammaVH, \one{\gammaVH} - \two{\gammaVH})$ are zero vectors, the so that those in $\text{Cov}\left(\bm{\std{\gammaVH}}, \bm{\std{(\one{\gammaVH} - \two{\gammaVH})}}\right)$.
Consequently, $\Cov{\bm{\std{\betaVH}}}{\bm{\std{\left(\betaAH - \betaBH'\right)}}}$ is a zero matrix, which satisfies the independence condition required in theorem 1.

\subsection*{Extension to GLM with linear projections of baseline covariates}

In this section we present an extension of the proof to accommodate GLM with linear projections of baseline covariates.
The columns of the $p \times q$ matrix $\bm{V}$ are the weights for the baseline covariates.
To perform a regression on the projected variables, $\XA$ and $\XB$ are first projected by $\bm{V}$.
As a result of the projection, the matrices $\X$, $\one{X}$ and $\two{X}$ will need to be re-defined.
The matrix $\X$ of the additive model, $\MA$, is now written as a linear projection of $\bm{D}$ in equation \eqref{eq:matD}:
\begin{equation}
\X = \bm{DU} = \left(\!
    \begin{array}{cccc}
      \bm{\grpA{X}V} &  \grpA{1} & \grpA{0} & \XCA\\
      \bm{\grpB{X}V} &  \grpB{0} & \grpB{1} & \XCB
    \end{array}
  \!\right) \text{ with }
\bm{U} = \left(\!
    \begin{array}{cc}
      \bm{V} &  \bm{0_{p \times (p_c+2)}}\\
      \bm{0_{(p_c+2) \times p }} & \bm{\mathrm{I}_{(p_c + 2) \times (p_c + 2)}}
    \end{array}
  \!\right),	
\label{eq:pcaX}
\end{equation}
where the dimensions of matrix $\bm{U}$ are $(p + p_c + 2) \times (p + p_c + 2)$.
The covariate matrix $\one{X}$ of the interaction model, $\model{I(1)}$, becomes
\begin{align}
\one{X} &= \one{D} \one{U} = \left(\!
    \begin{array}{ccccc}
      \bm{\grpA{X}V} &  \bm{0_{n_\mathrm{A} \times p}} & \grpA{1} & \grpA{0} & \XCA\\
      \bm{0_{n_\mathrm{B} \times p}} & \bm{\grpB{X}V} &  \grpB{0} & \grpB{1} & \XCB
    \end{array}
  \!\right) \nonumber\\
\one{U} &= \left(\!
    \begin{array}{ccc}
      \bm{V} &  \bm{0_{p \times p}} & \bm{0_{p \times (p_c+2)}}\\
      \bm{0_{p \times p}} & \bm{V} &   \bm{0_{p \times (p_c+2)}}\\
      \bm{0_{(p_c+2) \times p }} & \bm{0_{(p_c+2) \times p }} & \bm{\mathrm{I}_{(p_c + 2) \times (p_c + 2)}}
    \end{array}
  \!\right),	
\label{eq:pcaX1}
\end{align}
where $\one{D}$ is defined in equation \eqref{eq:matD1}. 
The square matrix $\one{U}$ has $2p + p_c + 2$ dimensions on each side.
Similarly, the covariate matrix $\two{X}$ of the interaction model, $\model{I(2)}$, is a linear projection of $\two{D}$ by $\two{U}$:
\begin{align}
\two{X} &= \two{D} \two{U} = \left(\!
   \begin{array}{ccccc}
      \bm{0_{n_\mathrm{A} \times p}} & \bm{\grpA{X}V} &  \grpA{0} & \grpA{1} & \XCA\\
      \bm{\grpB{X}V} &  \bm{0_{n_\mathrm{B} \times p}} & \grpB{1} & \grpB{0} & \XCB
    \end{array}
  \!\right) \nonumber\\
\two{U} &= \left(\!
    \begin{array}{ccc}
      \bm{0_{p \times p}} & \bm{V} &   \bm{0_{p \times (p_c+2)}}\\
      \bm{V} &  \bm{0_{p \times p}} &   \bm{0_{p \times (p_c+2)}}\\
      \bm{0_{(p_c+2) \times p }} & \bm{0_{(p_c+2) \times p }} & \bm{\mathrm{I}_{(p_c + 2) \times (p_c + 2)}}
    \end{array}
  \!\right),	
\label{eq:pcaX2}
\end{align}
where $\two{D}$ is defined in equation \eqref{eq:matD2} and the square matrix $\two{U}$ has the same dimensions as $\one{U}$.

Substituting the new definitions of $\X$, $\one{X}$ and $\two{X}$ described in equations \eqref{eq:pcaX}, \eqref{eq:pcaX1} and \eqref{eq:pcaX2} respectively into the covariance expression in equation \eqref{eq:covEx1}, and then following the same arguments and workings from there onwards, we will also arrive at equation \eqref{eq:covEx4}.
Therefore, the proof also holds for GLM with linear projections of baseline covariates.

In the case of PC-GLM, $q = p$ and the columns of $\bm{V}$ are the PC coefficients of $\bm{X_\mathrm{S}}$ defined in equation \eqref{eq:dataByGroup}.	
On the other hand, in the unblinded IRM approach, $q = 1$ and $\bm{V}$ is a vector containing the coefficients for the main effects of $\bm{X_\mathrm{S}}$ in the risk model.

\section{Web Appendix B: Supplementary analyses of the SEAQUAMAT trial data}\label{sec:supAnalysis}
\setcounter{figure}{0} 
In this section, the first part (\ref{sec:supAnalysisFull}) presents the analyses on the data contain all 1461 patients in the trial, where we apply single-stage screening of prognostic variables.
This and the primary analysis (in section 4.3) allow the comparison between single-stage and multi-stage screening.

In the second part (section \ref{sec:supAnalysis600}), we subsample the SEAQUAMAT data so the subsets only contain 600 patients.
We then apply the multi-staging screening as in section 4.3.

\subsection{Analyses with all patients recruited in the trial}\label{sec:supAnalysisFull}

\subsubsection{Interaction discovery with GLM after outcome-driven dimension reduction}
\label{subsec:esGLM}

\begin{figure}[ht]
  \centering
    \includegraphics[width=0.8\textwidth]{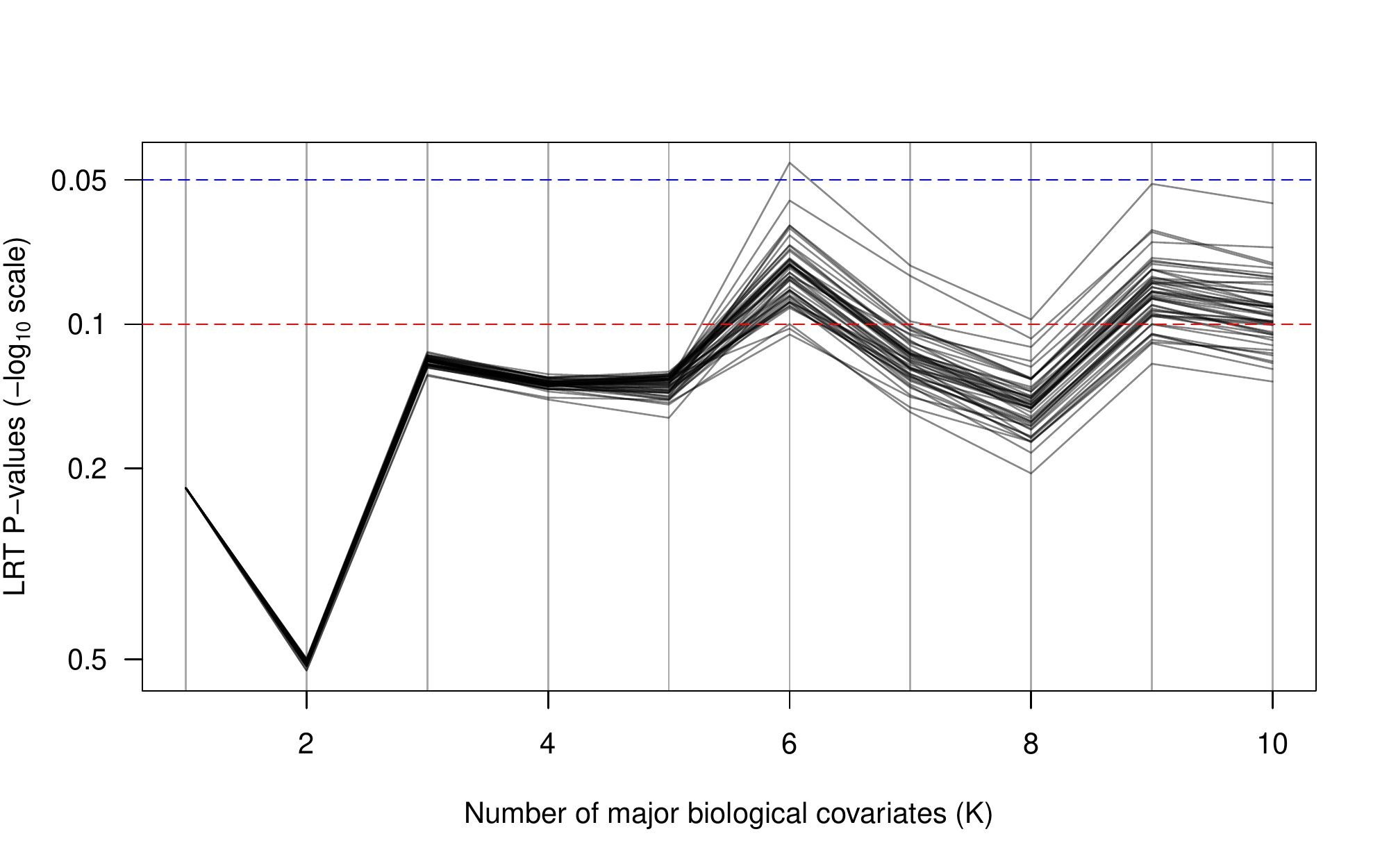}
  \caption{The LRT p-values versus degrees of freedom ($K$) across 50 imputed datasets of all 1461 patients recruited onto the SEAQUAMAT trial. 
  Here, $K$ the number of PCs of the variables selected by supervised variable selection at substage-II of Stage-1 of the interaction analysis. }
  \label{fig:fullMalariaGLMRegDFs}
\end{figure}

As the purpose of this demonstration is to showcase the properties of our approach, we iterate through $K$ values from 1 to 10.
The LRT p-values are presented in figure \ref{fig:fullMalariaGLMRegDFs}.

\begin{figure}[ht]
  \centering
    \includegraphics[width=0.7\textwidth]{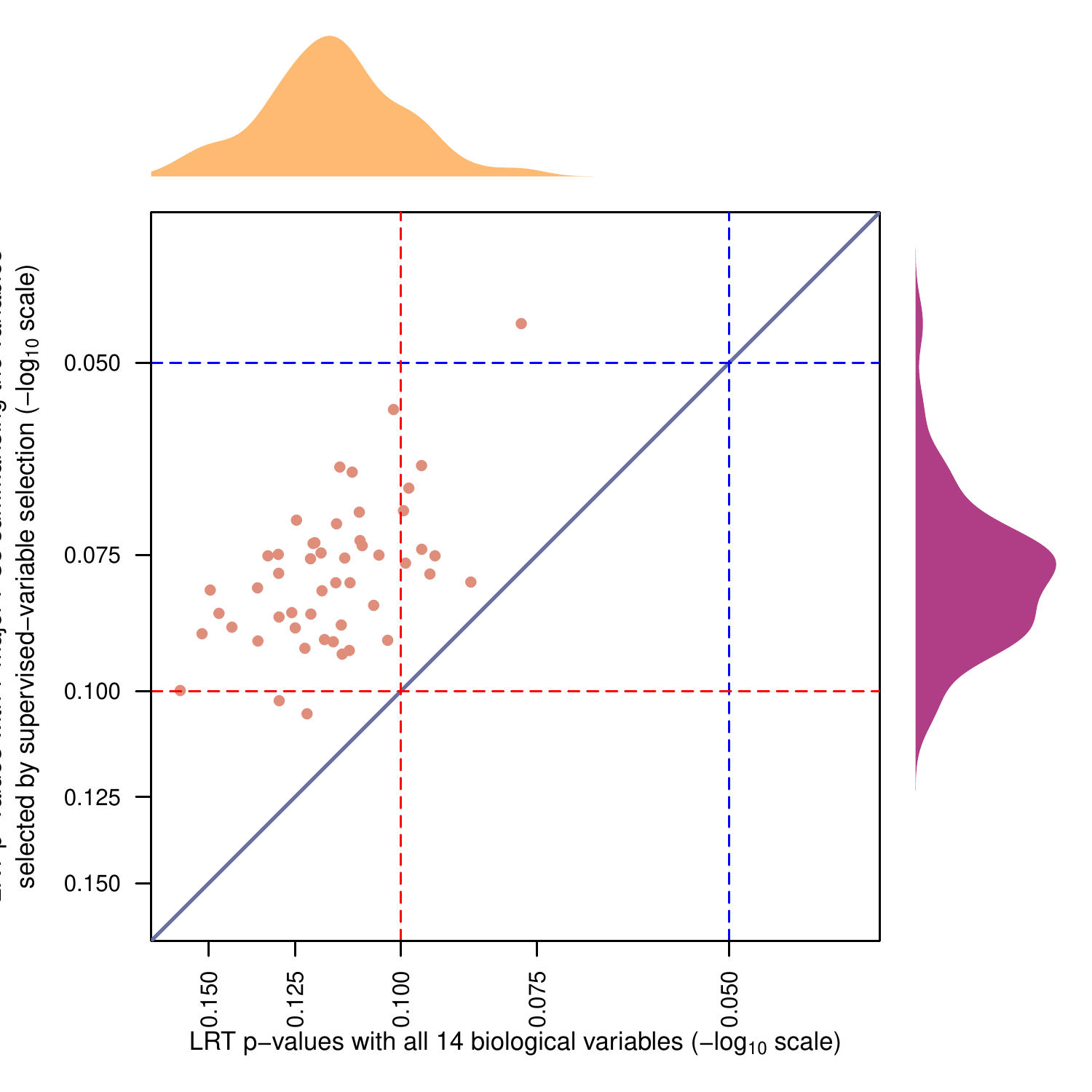}
  \caption{Comparison of LRT p-values in a GLM framework for the 1461-patient SEAQUAMAT data. 
  For each of the 50 imputed dataset, the LRT p-values for TEH are compared between the analysis with all 14 biological variables and that with 6 biological variables selected by GBM in Stage-1.
  The purple straight line goes through the origin and has a slope of 1.}
  \label{fig:malariaGLMLRTPvals}
\end{figure}

Across the different $K$ values, the evidence is the strongest when we test for the interaction between treatment and the set of 6 variables = \{coma, logged parasitaemia, age, heart rate, temperature, systolic blood pressure\}.
The interaction with this set provides a median p-value = 0.0787 (p-value range = [0.0460, 0.105], 3 s.f.) and p-values $<0.1$ are obtained from 48 of the 50 imputed datasets.
Moreover, figure \ref{fig:malariaGLMLRTPvals} illustrates that the evidence for TEH is stronger (LRT p-value decreases) for every imputed dataset when we reduce the variables tested from all 14 biological variables to the 6 variables selected by boosting.

As depicted in figure \ref{fig:fullMalariaGLMRegDFs}, further inclusion of the remaining biological candidates displayed some instability.
When the haematocrit variable is added to this set, the strength of evidence for interaction effect experiences a considerable drop with only three imputed datasets having p-value $<0.1$.
Subsequent inclusion of weight reduces to only one imputed dataset with p-value $<0.1$.
However, inclusion of jaundice pushes the number of datasets with p-value $<0.1$ up to 42 but adding on duration of fever leads to a reduction to 34 sets with p-value $<0.1$.

In summary, even though the strength of evidence is not as strong a the multi-stage analysis, it is observed that the power to detect interactions can be increased by single-stage screening of the main effects in the additive model.

\subsection{Analyses only with patients having staging data}\label{sec:supAnalysis600}

The full details of the analyses in this section can be found in R notebook S2: Empirical demonstration of interaction analysis on the data of the subset of patients from the SEAQUAMAT trial who have staging information.

Of the 1461 patients in the SEAQUAMAT trial, 854 have staging data of the malaria parasite lifecycle.
The data is in the form of percentages of the following stages: tiny ring, small ring, large ring, early troph, mid troph, late troph and gametocytes.
There are another two additional variables in this dataset, which are base deficit and indication of prior antimalarials treatment received.

The purpose of this section is to demonstrate that when the sample size of the SEAQUAMAT trial data is substantially reduced, that our framework can still achieve better power to detect TEH than na\"{i}vely including all the variables in the interaction test, even when we account for subsampling variation.
Section \ref{appB:fullAnalysis854} presents the problem encountered when including all biological variables in the interaction analysis on the 854 patients.
Section \ref{appB:subsampling} describes the subsampling of the SEAQUAMAT data, so that the problem in section \ref{appB:fullAnalysis854} will have a greater impact and that we can account for subsampling variation.
Subsequently, we apply our framework to each subset and present the results in section \ref{appB:pcGLM600}.
Here, we only consider multi-stage screening, which is our primary recommendation. 
In addition, we select the PC in Substage-II by its variance as that seems to exhibit more stable behaviour (section 4.3.2).

\subsubsection{Interaction discovery with GLM including all variables}\label{appB:fullAnalysis854}

\begin{figure}[ht]
  \centering
    \includegraphics[width=0.6\textwidth]{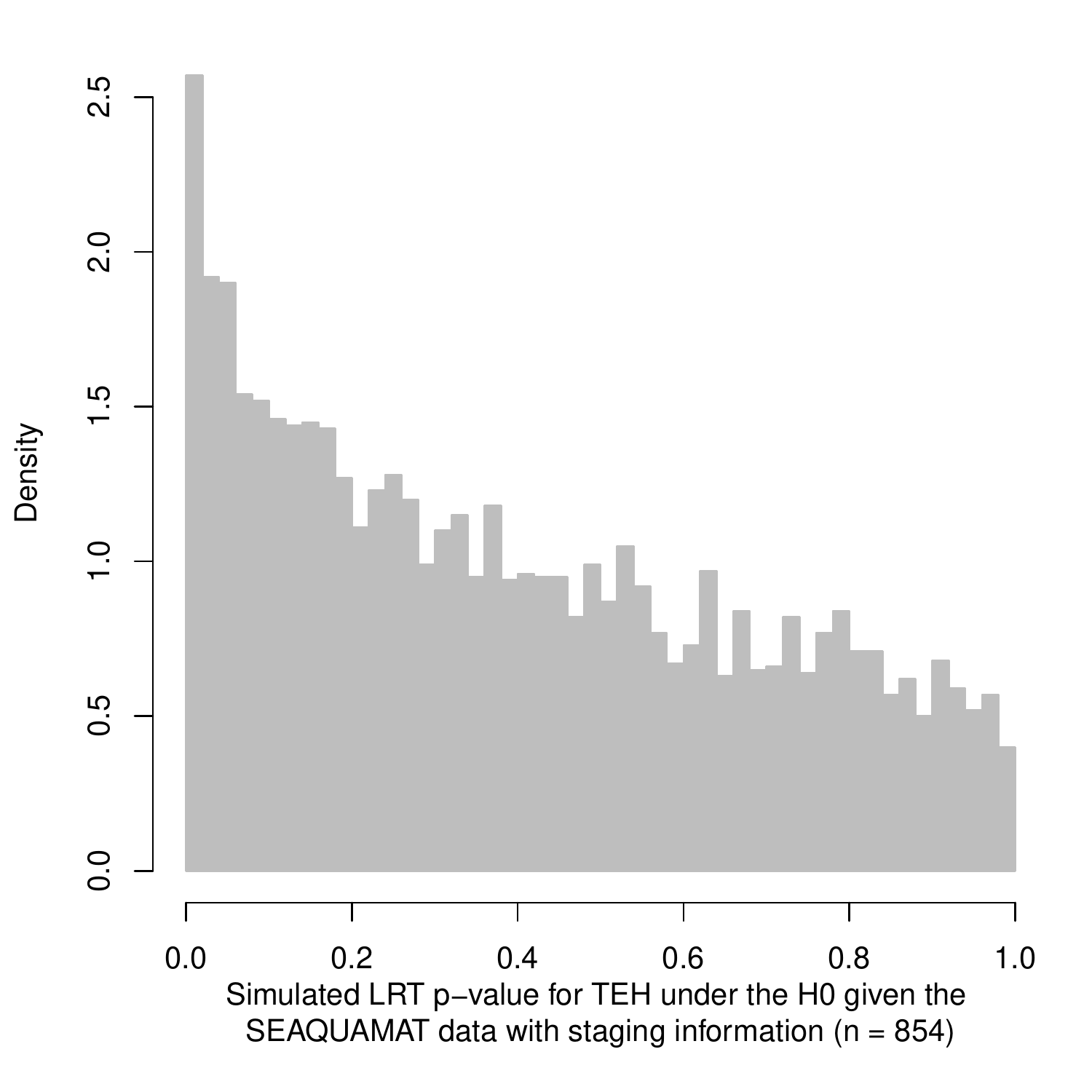}
  \caption{The simulated distribution of LRT P-values for TEH under the H0 given the SEAQUAMAT trial data with staging information (sample size = 854).}
  \label{fig:malariaStagingFullH0Pvals}
\end{figure}

After the inclusion of all the aforementioned additional variables, the degrees of freedom of the LRT test for interaction reaches 25, with 30 and 55 variables respectively in the full additive and the full interaction model. 
Given the results presented by \cite{sur:2017li}, it is reasonable to question the accuracy of the LRT p-value of the interaction test.
To investigate the bias, we simulate distribution of the LRT p-values under H0, and this simulated distribution is presented in figure \ref{fig:malariaStagingFullH0Pvals}, which clearly displays the underestimation of p-values.

Using the simulated distribution to correct for the observed bias, the full interaction test for TEH provides a p-value of 0.574 (3 s.f.), which is a borderline result that does not provides evidence against the global H0 of no interactions at the 5\% confidence level.

\subsubsection{Sub-sampling}\label{appB:subsampling}
While the result with the 854 patients is not statistically significant at the 5\% level, the LRT p-value is rather close to the 0.05 threshold. 
Therefore we subsample the data to 600 patients where there would be less power to detect TEH.

We have created 100 subsampled datasets.
As the SEAQUAMAT trial data is roughly balanced between the two treatment arms, the subsampled data subsampling is by stratified treatment arms.
For each subsample, we have simulated the LRT p-value distribution for H0 including all 25 biological interaction-candidates.
The LRT p-value for full treatment interaction of each dataset is corrected using the corresponding simulated H0 distribution.
We focus only on the 67 subsamples that have (corrected) LRT p-values greater than 0.1.

\subsubsection{Interaction discovery with PC-GLM after outcome-driven dimension reduction} \label{appB:pcGLM600}
Each of subsampled dataset is fitted with GBM with stumps. 
Relative influence is used to determine the priority of variable inclusion for testing the global TEH. 
For each subsample, there is still a fairly large number of variables with that have positive relative influence, where many of these variables have very low relative influence ($<$1), so they are likely to have little predictive information and we only select the variables with RI $>$1.  
Following this, PCA is applied to the variables selected for each subsampled dataset.

Because we have already performed the analysis on the full 1461-patient dataset with a range of $K$ values (section 4.3), we will not do the same here.
Instead we chose the 8 most variable PCs to include in the LRT to test whether they interact with the treatment.
We chose 8 because of the 67 subsets, the mininum number of variables with RI greater than 1 is 8, suggesting that when we account for subsampling variation, there are at least 8 variables that are informative of prognosis.
To protect from potential poor approximation by the $\chi^2$ distribution used in the LRT, the null distribution of the LRT p-values are simulated and used to correct the LRT p-values for the interactions between the 8 leading PCs and the treatment.

\begin{figure}[ht]
  \centering
    \includegraphics[width=0.6\textwidth]{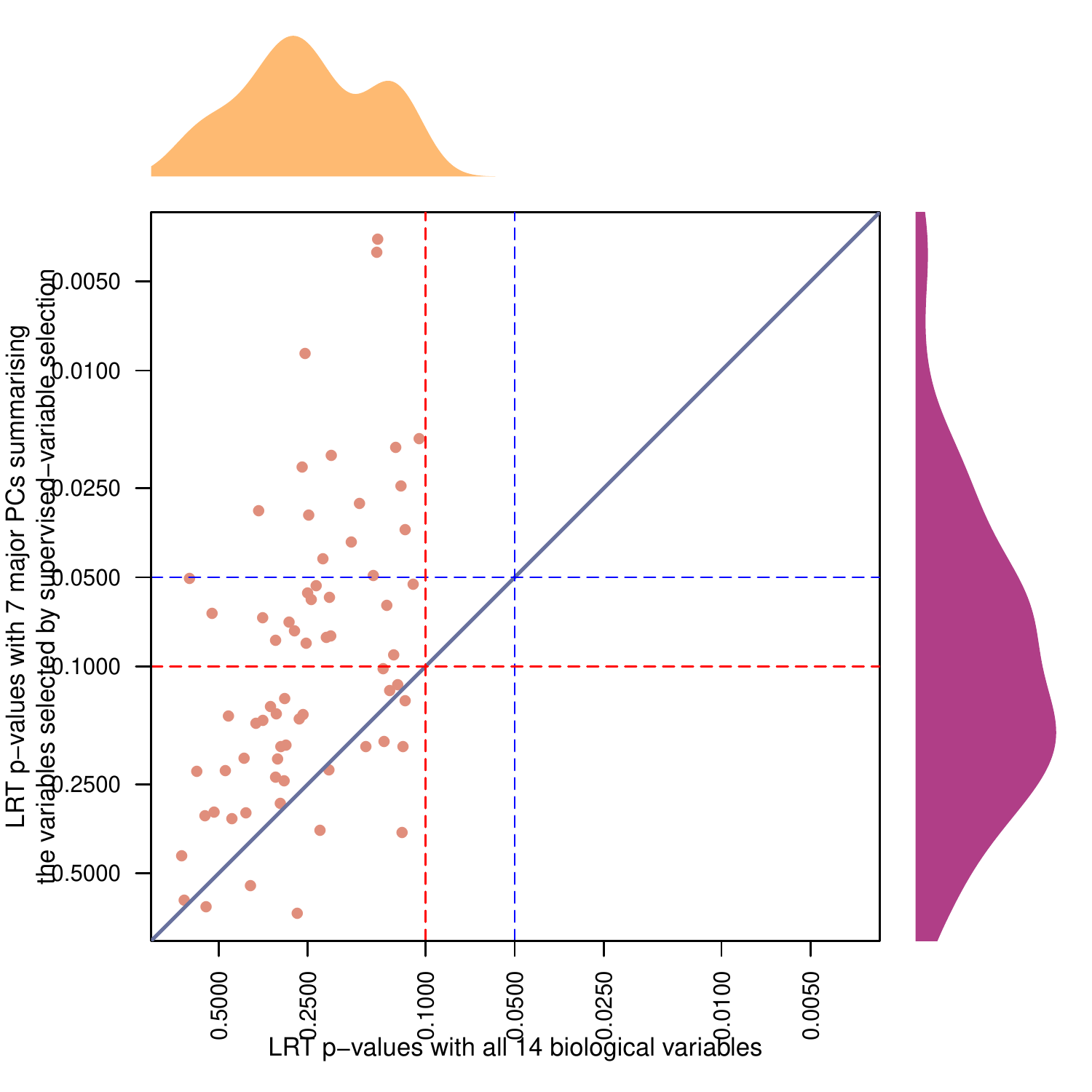}
  \caption{LRT P-values comparison in a PC-GLM framework for subsamples of SEAQUAMAT data with staging information (n=600). 
  For every of the 67 subsampled dataset, the LRT p-values for TEH are compared between the analysis with all 25 biological variables and that with 8 major PCs representing all the variables with relative influence $>1$ indicated by supervised-variable selection (GBM).
  The purple straight line goes through the origin and has a slope of 1.}
  \label{fig:malariaStagingPCAGLMLRTPvals}
\end{figure}

It is evident in figure \ref{fig:malariaStagingPCAGLMLRTPvals}, we improved power to detect TEH when we reduce the variables tested from 25 biological variables to 8 PCs representing the baseline covariates selected by GBM.
Fifty-seven of the 67 subsamples have improved in the strength of evidence for TEH. 
The strength of evidence for TEH varies considerably because subsampling patients will creates variation in the signal of TEH across the subsampled datasets.
Nevertheless, the general trend is that our framework can have better power to detect TEH than including all variables.

\end{document}